%% file: main.tex
\documentclass[letterpaper,twocolumn,10pt,table]{article}

\usepackage[draft]{hyperref}
\usepackage{usenix2019_v3}
\usepackage{url}
\usepackage{tikz}
\usepackage{amsmath}
\usepackage{amssymb}
\usepackage{listings}
\usepackage{xspace}
\usepackage{tabularx}
\usepackage{fontawesome}
\usepackage{graphicx}
\usepackage{caption}
\usepackage[utf8]{inputenc}
\usepackage{pgfplots}
\DeclareUnicodeCharacter{2212}{−}
\usepgfplotslibrary{groupplots,dateplot}
\usetikzlibrary{patterns,shapes.arrows}
\pgfplotsset{compat=newest}

\usepackage[draft,mode=multiuser]{fixme} 
\FXRegisterAuthor{gc}{angc}{GC}
\FXRegisterAuthor{aa}{anaa}{AA}
\FXRegisterAuthor{av}{anav}{AV}
\fxsetup{layout={marginclue,footnote}}
\fxusetheme{colorsig}

\newcommand{\toolname}{RevOK\xspace}

\newcommand{\Scantool}{Scanning system}
\newcommand{\scantool}{scanning system}
\newcommand{\nmap}{Nmap}
\newcommand{\tee}{TEE}
\newcommand{\ntools}{{78}}
\newcommand{\nvuln}{{36}}
\newcommand{\ntaint}{{67}}
\newcommand{\ipfingerprints}{{[{\it Anonymous}]}}

\begin{document}

\date{}

\title{Never Trust Your Victim: Weaponizing Vulnerabilities in Security Scanners}

\author{%
{\rm Andrea Valenza}\\
University of Genova\\
andrea.valenza@dibris.unige.it
\and
{\rm Gabriele Costa}\\
IMT School for Advanced Studies Lucca\\
gabriele.costa@imtlucca.it
\and
{\rm Alessandro Armando}\\
University of Genova\\
alessandro.armando@unige.it
}

\maketitle

\begin{abstract}
The first step of every attack is reconnaissance, i.e., to acquire information about the target.
A common belief is that there is almost no risk in scanning a target from a remote location.
In this paper we falsify this belief by showing that scanners are exposed to the same risks as
their targets.
Our methodology is based on a novel attacker model where the scan author becomes the victim of a counter-strike.
We developed a working prototype, called \toolname{}, and we applied it to \ntools{} \scantool{s}.
Out of them, \nvuln{} were found vulnerable to XSS.
Remarkably, \toolname{} also found a severe vulnerability
in Metasploit Pro, a mainstream penetration testing tool.
\end{abstract}


\input{introduction}

\input{background}

\input{threat-model}

\input{methodology}

\input{results}

\input{case-study}

\input{related}

\input{conclusion}

\bibliographystyle{plain}
\bibliography{main}

\flushcolsend 
\newpage
\begin{appendix}
\input{disclosure}
\end{appendix}
\end{document}

%% file: introduction.tex
\section{Introduction}\label{sec:introduction}

Performing a network scan of a target system is a surprisingly frequent operation.
There can be several agents behind a scan, e.g., attackers that gather technical information,
penetration testers searching for vulnerabilities, Internet users checking a suspicious address.
Often, when the motivations of the scan author are unknown, it is perceived by the target as a hostile operation.
However, scanning is so frequent that it is largely tolerated by the target.
Even from the perspective of the scanning agent, starting a scan seems not risky.
Although not completely stealthy, an attacker can be reasonably sure to remain anonymous by adopting basic precautions, such as proxies, virtual private networks and onion routing.  

Yet, expecting an acquiescent scan target is a mere assumption.
The \scantool{} may receive poisoned responses aiming to trigger vulnerabilities in the scanning host. 
Since most \scantool{s} generate an HTML report, scan authors can be exposed to attacks via their browser.
This occurs when the \scantool{} permits an unsanitized flow of information from the response to the user browser.
To illustrate, consider the following, minimal HTTP response. 
\begin{lstlisting}[basicstyle=\ttfamily,mathescape=true]
HTTP/1.1 200 OK
Server: nginx/1.17.0
$\ldots$
\end{lstlisting}
A naive \scantool{} might extract the value of the \verb|Server| field (namely, the string \verb|nginx/1.17.0| in the above example) and include it in the HTML report.
This implicitly allows the scan target to access the scan author's browser and inject malicious payloads.

In this paper we investigate this attack scenario. 
We start by defining an attacker model that precisely characterizes the threats informally introduced above.
To the best of our knowledge, this is the first time that such an attacker model is defined in literature.
Inspired by the attacker model, we define an effective methodology to discover cross-site scripting (XSS) vulnerabilities in the \scantool{s} and we implement a working prototype.
We applied our prototype to \ntools{} real-world \scantool{s}.
The results confirm our expectation: several (\nvuln{}) \scantool{s} convey attacks.
All of these vulnerabilities have been notified through a responsible disclosure process.

The most remarkable outcome of our activity is possibly an XSS vulnerability enabling remote code execution (RCE) in Rapid7 Metasploit Pro.
We show that the attack leads to the complete takeover of the scanning host.
Our notification prompted Rapid7 to undertake a wider assessment of their products based on our attacker model.

The main contributions of this paper are:
\begin{enumerate}
\item a novel attacker model affecting \scantool{s};
\item a testing methodology for finding vulnerabilities in \scantool{s};
\item \toolname{}, a prototype implementation of our testing methodology;
\item an analysis of the experimental results on \ntools{} real-world \scantool{s}, and;
\item three application scenarios highlighting the impact of our attacker model. 
\end{enumerate}

This paper is structured as follows.
Section~\ref{sec:background} recalls some preliminary notions.
Section~\ref{sec:threat-model} presents our attacker model.
We introduce our methodology in Section~\ref{sec:methodology}.
Our prototype and experimental results are given in Section~\ref{sec:implementation}.
Then, we present the three use cases in Section~\ref{sec:case-study}, while we survey on the related literature in Section~\ref{sec:related}.
Finally, Section~\ref{sec:conclusion} concludes the paper.


%% file: background.tex
\section{Background}\label{sec:background}

In this section we recall some preliminary notions necessary to correctly understand our methodology. 

\subsection{\Scantool{s}}
\label{sec:scantools}

A \scantool{} is a piece of software that $(i)$ stimulates a target through network requests, $(ii)$ collects the responses, and $(iii)$ compiles a report.
Security analysts often use \scantool{s} for technical information gathering~\cite{mitre-techinfogather}.
\Scantool{s} used for this purpose are called \emph{security scanners}.
Our definition encompasses a wide range of systems,
from complex vulnerability scanners
to simple ping utilities. 

\begin{figure}
    \includegraphics[width=\columnwidth]{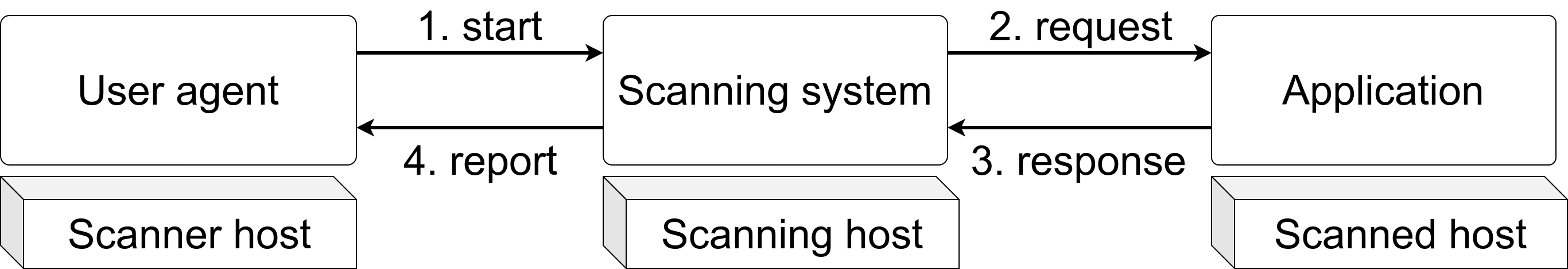}
    \caption{Abstract architecture of a \scantool{}.}\label{fig:scan}\label{fig:std-scan}
\end{figure}

Figure~\ref{fig:scan} shows the key actors involved in a scan process.
Human analysts use a user agent, e.g., a web browser,
to select a target, possibly setting some parameters,
and start the scan (1. start).
Then, the \scantool{} crafts and sends request messages to the target (2. request).
The \scantool{} parses the received response messages
(3. response), extracts the relevant information and provides the analyst with the scan result (4. report).
Finally, the analysts inspect the report via their user agent.

Whenever a \scantool{} runs on a separate, remote scanning host,
we say that it is provided \emph{as-a-service}.
Instead, when the scanner and scanning hosts coincide,
we say that the \scantool{} is \emph{on-premise}.

A popular, command line \scantool{} is \nmap{}~\cite{nmap-tool}.
To start a scan, the analyst runs a command from the command line, such as
\begin{center}
\verb|nmap -sV 172.16.1.26 -oX report.xml|
\end{center}
Then, \nmap{} scans the target (\texttt{172.16.1.26}) with requests aimed at identifying its active services (\verb|-sV|).
By default, \nmap{} sends requests to 1,000 frequently used TCP ports and collects responses from the services running on the target.
The result of the scan is then saved (\verb|-oX|) on \texttt{report.xml}.
Interestingly, some web applications, e.g., \nmap{} Online~\cite{nmap-online}, provide the functionalities of \nmap{} as-a-service.

\Scantool{s} are often components of larger, more complex systems, sometimes providing a browser-based GUI.
For instance, Rapid7 Metasploit Pro is a full-fledged penetration testing software.
Among its many functionalities, Metasploit Pro also performs automated information gathering, even including vulnerability scanning.
The reporting system of Metasploit Pro is based on an interactive Web UI used to browse the report.


\subsection{Taint analysis}
\label{sec:taintanalysis}

Taint analysis~\cite{Schwartz10taint} refers to the techniques used to detect how the information flows within a program.
Programs read inputs from some sources, e.g., files, and write outputs to some destinations, e.g., network connections.
For instance, taint analysis is used to understand whether an attacker can force a program to generate undesired/illegal outputs by manipulating some of its inputs.
A \emph{tainted flow} occurs when (part of) the input provided by the attacker is included in the (tainted) output of the program.
In this way, the attacker controls the tainted output which can be used to inject malicious payloads to the output recipient.

\subsection{Cross-site scripting}
\label{sec:xss}

Cross-site scripting (XSS) is a major attack vector for the web, stably in the OWASP Top 10 vulnerabilities~\cite{OWASP17topten} since its initial release in 2003.
Briefly, an XSS attack occurs when the attacker injects a third-party web page with an executable script, e.g., a JavaScript fragment.
The script is then executed by the victim's browser.
The simplest payload for showing that a web application suffers from an XSS vulnerability is
\begin{center}
\verb|<script>alert(1)</script>|
\end{center}
that causes the browser to display an alert window.
This payload is often used as a proof-of-concept (PoC) to safely prove
the existence of an XSS vulnerability.

There are several variants to XSS.
Among them, \emph{stored} XSS has highly disruptive potential.
An attacker can exploit a stored XSS on a vulnerable web application to permanently save the malicious payload on the server.
In this way, the attack is directly conveyed by the server that delivers the injected web page to all of its clients.
Another variant is \emph{blind} XSS, in which the attacker cannot observe the injected page.
For this reason, blind XSS relies on a few payloads, each adapting to multiple HTML contexts.
These payloads are called \emph{polyglots}.
A remarkable example is the polyglot presented in~\cite{elsobky20ultimatexss} which adapts to at least 26 different contexts.



%% file: threat-model.tex
\section{Attacker model}\label{sec:threat-model}

\begin{figure}
    \includegraphics[width=\columnwidth]{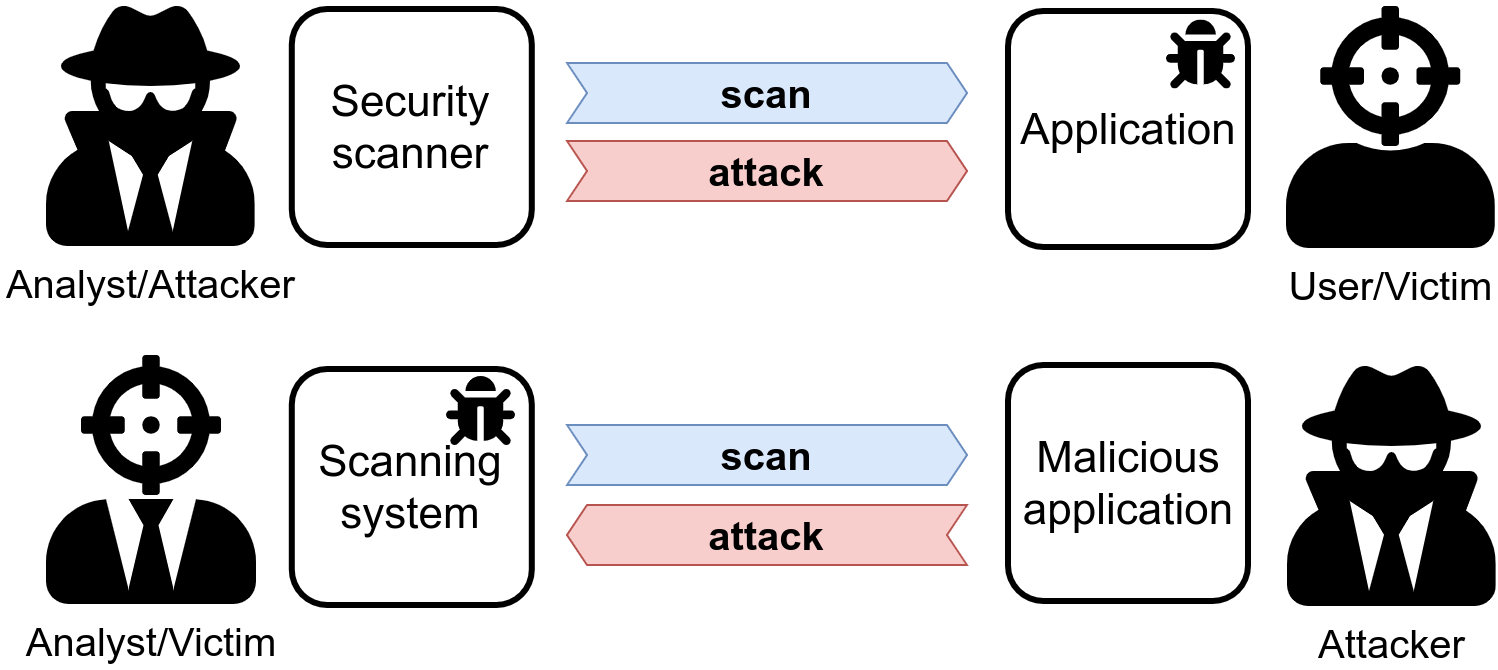}
    \caption{Comparison between attacker models.}
    \label{fig:attacker-model}
\end{figure}

The idea behind our attacker model is sketched in Figure~\ref{fig:attacker-model} (bottom), where we compare it with a
traditional web security attacker model (top).
Typically, attackers use a security scanner to gather technical
information about a target application.
If the application suffers from some vulnerabilities, attackers
can exploit them to deliver an attack towards their victims,
e.g., the application users.
On the contrary, in our attacker model attackers use malicious
applications to attack the author of a scan, e.g.,
a security analyst.

Here are the two novelties of our attacker model.
\begin{enumerate}
	\item Attacks are delivered through HTTP \emph{responses} instead of \emph{requests}.
	\item Attackers exploit the vulnerabilities of \scantool{s} to
	strike their victims, i.e., the scan initiator.
\end{enumerate}

Below, we detail the attacker's goal and capabilities.

\paragraph{Attacker goal.}

The objective of the attacker is to directly strike the analyst.
To do so, the attacker exploits the vulnerabilities of the target \scantool{} and its reporting system to hit the analyst user agent.
In this work, we assume that the user agent is a web browser.
This assumption covers every as-a-service \scantool{}, as well as
many on-premise ones, which generate HTML reports.
As a consequence, here we focus on XSS which is a major attack vector for web browsers.
As usual in XSS, the attacker succeeds when the victim's browser executes a piece of attacker-provided code, e.g., JavaScript.

\paragraph{Attacker capabilities.}

First, we state that the attacker has adequate resources
to detect vulnerabilities in \scantool{s} before deploying
the malicious application.
However, the attacker capabilities do not include the possibility
of observing the internal logic of the \scantool{}.
That is, our attacker operates in black-box mode.

Secondly, our attacker has complete control over the malicious application, e.g., the attacker owns the scanned host.
However, we do not assume that the attacker can force the victim to initiate the scanning process.

%% file: methodology.tex
\section{Testing methodology}\label{sec:methodology}

In this section, we define a vulnerability detection methodology
based on our attacker model.

\subsection{Test execution environment}
\label{sec:tee}

Our methodology relies on a \emph{test execution environment}
(\tee{}) to automatically detect vulnerabilities in \scantool{s}.
In particular, a \emph{test driver} simulates the user agent of the security analyst, while a \emph{test stub}
simulates the scanned application.
Our \tee{} can $(i)$ start a new scan, $(ii)$ receives the requests of the \scantool{}, $(iii)$ craft the responses of the target application, and $(iv)$ access the report of the \scantool{}.
Intuitively, the \tee{} replicates the configuration of Figure~\ref{fig:scan}.
In this configuration, the test driver is executed by the scanner host, and the test stub runs on the scanned host.
In general, the test driver is customized for each \scantool{} under testing.
For instance, it may consist of a Selenium-enabled~\cite{Gojare15selenium} browser stimulating the web UI of the \scantool{}.

Both the test driver and the test stub consist of some submodules.
These submodules are responsible for implementing the two phases
described below.

\subsection{Phase 1: tainted flows enumeration}\label{sec:find-tainted-flows}

\begin{figure}
    \includegraphics[width=\columnwidth]{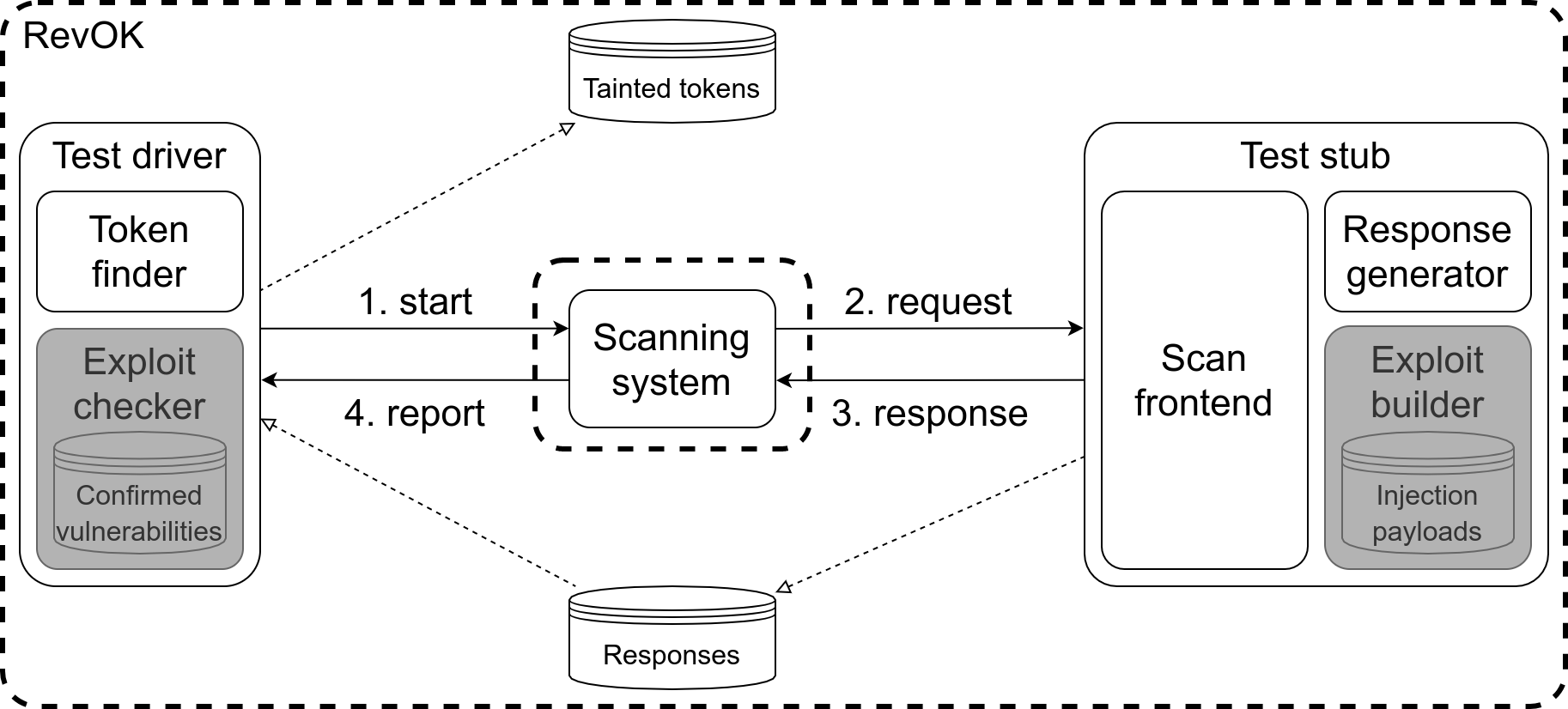}
    \caption{Phase 1 -- find tainted flows.}
    \label{fig:tainted}
\end{figure}

The first phase aims at detecting the existing tainted destinations in the report generated by the \scantool{}.
The process is depicted in Figure~\ref{fig:tainted}.
Initially, the test driver asks the \scantool{} to perform a scan of the test stub.
The scan logic is not exposed by the \scantool{} and, thus, it is opaque from our perspective.
Nevertheless, it generates some requests toward the test stub.
Each request is received by the \emph{scan frontend} and
dispatched to the \emph{response generator}, which crafts the
response. 

The response generation process requires special attention.
One might think that a single, general-purpose response is sufficient.
However, some \scantool{s} process the responses in non-trivial ways.
For instance, they may abort the scan if a malformed or suspicious response is received.
For this reason, we proceed as follows.
First, we generate a response template, i.e., an HTTP response containing variables, denoted by $t$.
Response templates are generated from a fuzzer through a \emph{probabilistic context-free grammar} (PCFG).
A PCFG is a tuple $(N, \Sigma, R, S, P)$, where
$G = (N, \Sigma, R, S)$ is a context-free grammar such that $N$ is the set of non-terminal symbols, $\Sigma$ is the set of terminal symbols, $R$ are the production rules and $S$ is the starting symbol.
The additional component of the PCFG, namely $P : R \rightarrow [0,1]$, associates each rule in $R$ with a probability, i.e., the probability to be selected by the fuzzer generating a string of $G$.
Additionally, we require that $P$ is a probability distribution over each non-terminal $\alpha$, in symbols
$$\forall \alpha \in N . \sum\limits_{(\alpha \,\mathtt{\mapsto}\,\beta)\, \in\, R} P(\alpha\, \mathtt{\mapsto}\,\beta) = 1$$
In the following, we write $\alpha \mapsto_{p} \beta$ for $P(\alpha \mapsto \beta) = p$ and $\alpha \mapsto_{p_1} \beta_1 \vert_{p_2} \ldots \vert_{p_n} \beta_n$ for $\alpha \mapsto_{p_1} \, \beta_1, \ldots, \alpha \mapsto_{p_n} \, \beta_n, \alpha \mapsto_{p_e}$ \verb|""| (where \verb|""| is the empty string).

The probability values appearing in our PCFG are assigned according to the results presented in~\cite{lavrenovs2018securityheaders, lavrenovs2019responseheaders}.
There, the authors provide a statistical analysis of the frequency of real response headers as well as a list of information-revealing ones.
Such headers are thus likely to be reported by a \scantool{}.
Finally, when the frequency of a field is not given (e.g., for variables), we apply the uniform distribution.

\begin{figure}[thb]
\begin{lstlisting}[basicstyle=\small\ttfamily,mathescape]
Resp $\mapsto_1$ Vers Stat Head Body 
Vers $\mapsto_{0.5}$  "HTTP/1.0" |$_{0.5}$  "HTTP/1.1"
Stat $\mapsto_{0.554}$ Succ |$_{0.427}$ Redr |$_{0.013}$ ClEr |$_{0.006}$ SvEr
Succ $\mapsto_{0.5}$ "200 OK" |$_{0.5}$ "200" $t$
Redr $\mapsto_{0.386}$ "301 Moved Permanently" |$_{0.386}$ "301" $t$ 
      |$_{0.114}$ "302 Found" |$_{0.114}$ "302" $t$
ClEr $\mapsto_{0.26}$ "403 Forbidden" |$_{0.26}$ "403" $t$ 
      |$_{0.24}$ "404 Not Found" |$_{0.24}$ "404" $t$
SvEr $\mapsto_{0.5}$ "500 Internal Server Error" |$_{0.5}$ "500" $t$
Head $\mapsto_1$ Serv PwBy Locn SetC CntT AspV MvcV Varn
         $\hookrightarrow$ StTS CnSP XSSP FrOp
Serv $\mapsto_{0.475}$ "Server:" $t$ |$_{0.475}$ "Server:" SrvT $t$
PwBy $\mapsto_{0.24}$ "X-Powered-By: php" 
      |$_{0.24}$ "X-Powered-By:" $t$
Locn $\mapsto_{0.315}$ "Location:" Link $|_{0.315}$ "Location:" $t$
Link $\mapsto_{0.516}$ "https://" $t$ 
      |$_{0.167}$ "http://" $t$ ":8899"
      |$_{0.135}$ "http://" $t$ ":8090"
      |$_{0.065}$ "http://" $t$ "/login.lp"
      |$_{0.059}$ "/nocookies.html"
      |$_{0.058}$ "cookiechecker?uri=/"
SetC $\mapsto_{0.175}$ "Set-Cookie:" Ckie
Ckie $\mapsto_{0.471}$ "__cfduid=" $t$ |$_{0.394}$ "PHPSESSID=" $t$
      |$_{0.087}$ "ASP.NET Session=" $t$
      |$_{0.048}$ "JSESSIONID=" $t$
CntT $\mapsto_{0.07}$ "X-Content-Type-Options: nosniff"
      |$_{0.07}$ "X-Content-Type-Options:" $t$
AspV $\mapsto_{0.5}$ "X-AspNet-Version:" $t$
MvcV $\mapsto_{0.5}$ "X-AspNetMvc-Version:" $t$
Varn $\mapsto_{0.5}$ "X-Varnish:" $t$
StTS $\mapsto_{0.5}$ "Strict-Transport-Security:" STSA
STSA $\mapsto_{0.111}$ "max-age=" $\mathbb{N}^+$ 
      |$_{0.111}$ "max-age=" $t$
      |$_{0.111}$ "max-age=" $\mathbb{N}^+$ "; preload"
      |$_{0.111}$ "max-age=" $t$ "; preload"
\end{lstlisting}
\caption{Response template grammar (excerpt).}
\label{fig:grammar}
\end{figure}

An excerpt of our PCFG is given in Figure~\ref{fig:grammar}.
For the sake of presentation, here we omit some of the rules and we refer the interested reader to the project web site\footnote{\url{https://github.com/AvalZ/RevOK}}.
The grammar defines the structure of a generic HTTP response (\verb|Resp|) made of a version (\verb|Vers|), a status (\verb|Stat|), a list of headers (\verb|Head|), and a body (\verb|Body|).
Variables $t$ are all fresh and they can appear in several parts of the generated response template.
In particular, variables can be located in status messages (i.e., \verb|Succ|, \verb|Redr|, \verb|ClEr| and \verb|SvEr|), header fields (i.e., \verb|Serv|, \verb|PwBy|, \verb|Locn|, \verb|SetC|, \verb|CntT|, \verb|AspV|, \verb|MvcV|, \verb|Varn|, \verb|StTS|, \verb|CnSP|, \verb|XSSP| and \verb|FrOp|) and body.
For instance, a field can be \verb|Server: nginx/|$t$, where \verb|nginx/| is a server type (\verb|SrvT|, omitted for brevity).

The response template is then populated by replacing each variable with a \emph{token}.
A token is a unique sequence of characters that is both \emph{recognizable}, i.e., it has a negligible probability of appearing by chance, and \emph{uninterpreted}, i.e., the browser treats it as plain text, when appearing in an HTML document.
All tokens are mapped to the responses containing them.
Responses are stored in a database.
Finally, the test driver matches the tokens appearing in the responses database with those occurring in the scan report.
Such tokens are evidence that there are tainted flows in the internal logic of the \scantool{}.
Tokens mark the source and the sink of a flow in the response and report, respectively. 
All these tokens are stored in the tainted tokens database.

\subsection{Phase 2: vulnerable flows identification}\label{sec:find-vuln-flows}

\begin{figure}
    \includegraphics[width=\columnwidth]{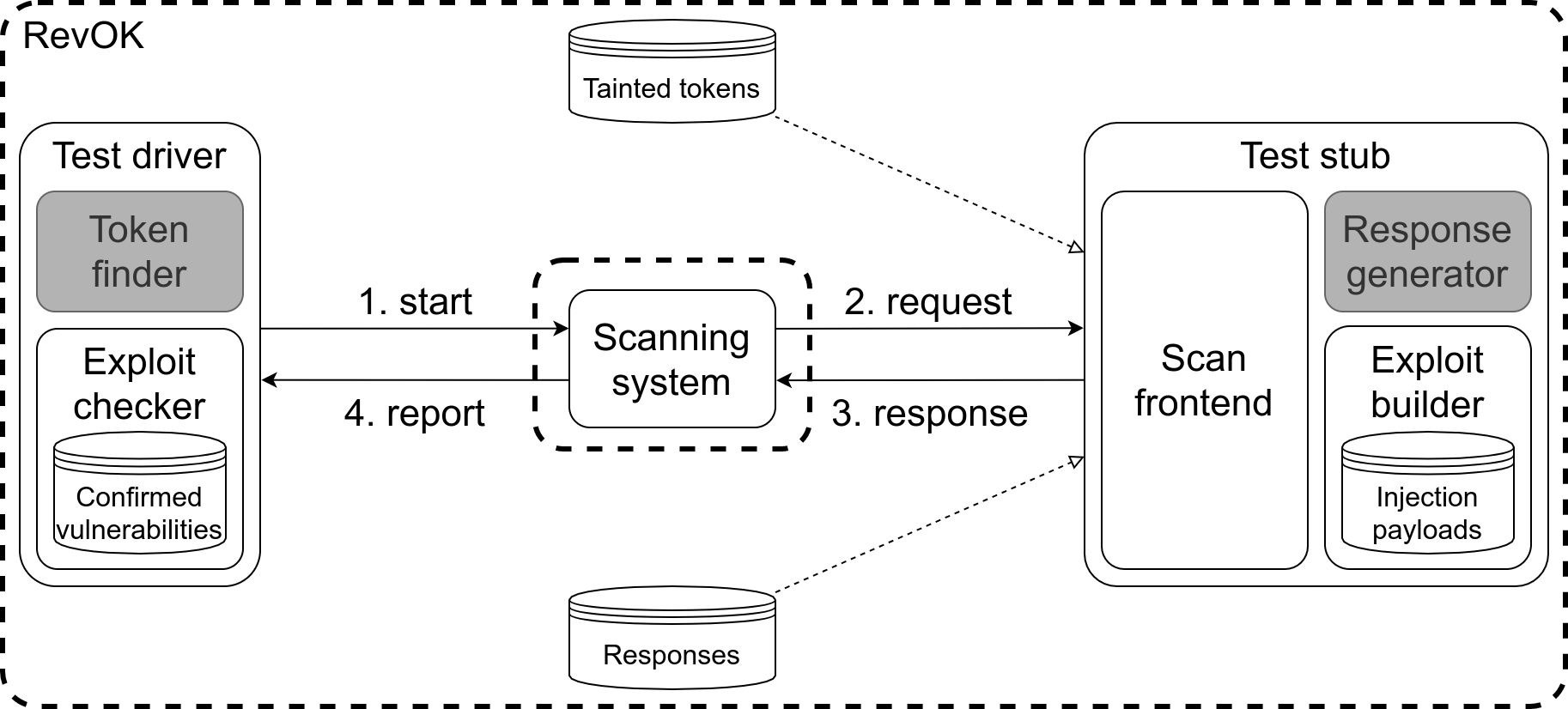}
    \caption{Phase 2 -- find vulnerable flows.}\label{fig:vulnerable}
\end{figure}

The second phase aims to confirm which tainted flows are actually vulnerable.
We use PoC exploits to confirm the vulnerability.
The workflow is depicted in Figure~\ref{fig:vulnerable}.
As for the first phase, the test driver launches a scan of the test stub.
When the test stub receives the requests, the exploit builder extracts a response from the responses database.
Then, the response is injected with a PoC exploit.
More precisely, a tainted token is selected among those generated during Phase 1.
The tainted token in the response is replaced 
with a payload taken from a predefined injection payload database. 
In general, a vulnerability is confirmed by the test driver according to predefined, exploit-dependent heuristics.
Although tainted flows can be subject to different types of vulnerabilities, as discussed in Section~\ref{sec:threat-model}, we focus on XSS.
Thus, the heuristics implemented by the exploit checker
consists of recognizing a vulnerable flow when an alert window is spawned by the corresponding, tainted flow.
Finally, the exploit checker stores the vulnerable flows
in the confirmed vulnerabilities database.

The definition of injection payload is non-trivial.
Since our \tee{} applies to both on-premise and as-a-service
\scantool{s}, some issues must be considered.
The first issue is testing performances.
As a matter of fact, \scantool{s} can take a considerable amount of
time to perform a single scan.
Moreover, as-a-service \scantool{s} should not be flooded by requests
to avoid degradation of the quality of service.
For these reasons, we aim to limit the number of payloads to check.

As discussed in Section~\ref{sec:xss}, polyglots allow us to test
multiple contexts with a single payload.
In this way, we increase the success probability
of each payload and, thus, we reduce the overall number of tests.

In principle, we might resort to the polyglot of~\cite{elsobky20ultimatexss}, which escapes 26 contexts.
However, its length (144 characters) is not adequate since many \scantool{s} shorten long strings when compiling their reports, so preventing the exploit from taking place.
To avoid this issue, we opted for polyglots such as
\verb|"'/><img src='x' onerror='alert(1)'/>|.
This is rendered by the browser when appearing inside both an HTML tag and an HTML attribute. 
The reason is that the initial \verb|"| and \verb|'| allow the payload to escape from quoted attributes.

Furthermore, delivering the JavaScript payload in \verb|onerror| has two advantages.
First, it circumvents basic input filtering methods, e.g., blacklisting of the \verb|script| string.
Secondly, our payload applies to both static and dynamic reports.
More precisely, a static report consists of HTML pages that are created by the \scantool{} and subsequently loaded by the analyst's browser.
Instead, a dynamic report is loaded by the browser and updated by the \scantool{} during the scan process.
The HTML5 standard specification~\cite[\S\,8.4.3]{whatwg20html5} clearly states that browsers can skip the execution of dynamically loaded scripts.
For this reason, our payload binds the script execution to an \emph{error} event that we trigger using a broken image link (i.e., \verb|src='x'|).
A concrete example of this scenario is discussed in Section~\ref{sec:metasploit}.


%% file: results.tex
\section{Implementation and results}
\label{sec:implementation}

In this section, we present our prototype \toolname{}.
We used it to carry out an experimental assessment that we discuss in Section~\ref{sec:results}.
 
\subsection{\toolname{}}
\label{sec:prototype}

Our prototype consists of two modules: the test driver and
the test stub.
We detail them below.

\paragraph{Test driver}
A dedicated test driver is used for each \scantool{}.
The test driver $(i)$ triggers a scan against the test stub, $(ii)$ saves the report in HTML format and $(iii)$ processes the report to detect tainted and vulnerable flows (in Phase 1 and 2, respectively).
While $(iii)$ is the same for all the \scantool{s}, $(i)$ and $(ii)$  may vary.

In general, the implementation of $(i)$ and $(ii)$ belongs to two categories depending on whether the \scantool{} has a programmable interface or only a GUI.
When a programmable interface is available, we implement a Python 3 client application.
For instance, we use the native \emph{os} Python module to launch \nmap{} so that its report is saved in a specific location (as describe in Section~\ref{sec:background}).
Similarly, we use the \emph{requests} Python library\footnote{\url{https://requests.readthedocs.io}} to invoke the REST APIs provided by a \scantool{} and save the returned HTML report.
Instead, when the \scantool{} only supports GUI-based interactions, we resort to GUI automation.
In particular, we use the Selenium Python library\footnote{\url{https://selenium-python.readthedocs.io/}} for browser-based GUIs and \emph{PyAutoGUI}\footnote{\url{https://pyautogui.readthedocs.io}} for desktop GUIs.
In the case of GUI automation, the test driver repeats a sequence of operations recorded during a manual test.

Finally, for the report processing step $(iii)$ we distinguish between two operations.
The tainted flow detection trivially searches the report for the injected tokens provided by the response generator (see below).
Instead, vulnerable flows are confirmed by checking the presence of alert windows through the Selenium function \texttt{switch\_to\_alert()}.

\paragraph{Test stub}
For the response generator, we implemented the PCFG grammar fuzzer detailed in Section~\ref{sec:find-tainted-flows} in Python.
Tokens are represented by randomly-generated Universally Unique Identifiers~\cite{rfc05uuid} (UUID).
A UUID consists of 32 hexadecimal characters organized in 5 groups that are separated by the \verb|-| symbol.
En example UUID is \verb|018d54ae-b0d3-4e89-aa32-6f5106e00683|.
As required in Section~\ref{sec:find-tainted-flows}, UUIDs are both recognizable (as collisions are extremely unlikely to happen) and uninterpreted (as they contain no HTML special characters).

On the other hand, starting from a response, the exploit builder replaces a given UUID with an injection payload.
Payloads are taken for a predefined list of selected polyglots, as discussed in Section~\ref{sec:find-vuln-flows}.

\subsection{Selection criteria}\label{sec:selection-criteria}

We applied our prototype implementation to \ntools{} \scantool{s}.
The full list of \scantool{s}, together with our experimental
results (see Section~\ref{sec:results}), is given in Table~\ref{tab:results}.
There, we use \faGlobe{} and \faArrowCircleODown{} to distinguish
between as-a-service and on-premise \scantool{s}, respectively.

For our experiments, we searched for \scantool{s} included in several
categories.
In particular, we considered security scanners, server fingerprinting tools, search engine optimization (SEO) tools,
redirect checkers, and more.
From these, we removed \scantool{s} belonging to
the following categories.
\begin{itemize}
	\item Abandonware, i.e., on-premise \scantool{s} that were not maintained in the last 5 years.
	\item Paywalled, i.e., \scantool{s} that are not free and have no trial version.
	\item Scheduled, i.e., as-a-service \scantool{s} that only perform periodic scans, not controlled by the analyst.
\end{itemize}

\subsection{Results}
\label{sec:results}

We applied \toolname{} to the \scantool{s} of Table~\ref{tab:results}.
For each \scantool{}, we used \toolname{} to execute 10 scan rounds
and we listed all the detected tainted and vulnerable flows.
As a result, we discovered that \ntaint{} \scantool{s} have tainted flows and, among them, \nvuln{} are vulnerable to XSS.

\newcolumntype{s}{>{\hsize=.05\hsize}X}

\begin{table*}[t]
	\caption{Experimental results for the considered \scantool{s}.
		($\dagger$ requested to stay anonymous.) \\ T is the number of tainted flows, V is the number of vulnerable flows, M indicates if a human analyst found a vulnerability. }\label{tab:results}
	\small
	\rowcolors{2}{white}{gray!10}
	\begin{tabularx}{\textwidth}{|sXrrs|sXrrs|sXrrs|}
	\hline

		& \textbf{Name} & \textbf{T} & \textbf{V} & \textbf{M} &  & \textbf{Name} & \textbf{T} & \textbf{V} & \textbf{M} & & \textbf{Name} & \textbf{T} & \textbf{V} & \textbf{M} \\
	  \hline
	  \faGlobe & AddMe                     & 11 & 11 & \checkmark{} & \faGlobe & InternetOfficer       & 2  & 1  & \checkmark{} & \faGlobe & Security Headers      & 13 & -  &   \\
	  \faGlobe & AdResults                 & 14 & -  &   & \faGlobe & \ipfingerprints{}$^\dagger$ & 11 & 1  & \checkmark{} & \faGlobe & SEO Review Tools      & -  &  -  &   \\
	  \faArrowCircleODown & Arachni                   & 14 & -  &   & \faGlobe & iplocation.net        & -  & -  &   & \faGlobe & SeoBook               & 12 & 11 & \checkmark{}\\
	  \faGlobe & AUKSEO                    & -  & -  &   & \faGlobe & IPv6 Scanner          & -  & -  &   & \faGlobe & SERP-Eye              & -  & -  &   \\
	  \faGlobe & BeautifyTools             & 13 & -  &   & \faGlobe & itEXPERsT             & -  & -  &   & \faGlobe & Server Headers        & 13 & 12 & \checkmark{}\\
	  \faGlobe & BrowserSPY                & 9  & -  &   & \faArrowCircleODown & IVRE                  & 2  & -  &   & \faGlobe & Site 24x7             & 13 & 13 & \checkmark{}\\
	  \faGlobe & CheckHost                 & 1  & -  &   & \faGlobe & JoydeepDeb            & 13 & 13 & \checkmark{} & \faGlobe & SQLMap Scanner        & 1  & 1  & \checkmark{}\\
	  \faGlobe & CheckMyHeaders.com        & 1  & -  &   & \faGlobe & JSON Formatter        & 13 & 13 & \checkmark{} & \faGlobe & SSL Certificate Tools & 12 & -  &   \\
	  \faGlobe & CheckSERP                 & 11 & -  &   & \faGlobe & LucasZ ZeleznY        & 2  & 1  & \checkmark{} & \faGlobe & StepForth             & 12 & 11 & \checkmark{}\\
	  \faGlobe & CheckShortURL             & 1  & 1  & \checkmark{} & \faArrowCircleODown & Metasploit Pro        & 11 & 3  & \checkmark{} & \faGlobe & StraightNorth         & -  & -  &   \\
	  \faGlobe & Cloxy Tools               & 11 & -  &   & \faGlobe & Monitor Backlinks     & 12 & -  &   & \faGlobe & SubnetOnline          & 14 & 13 & \checkmark{}\\
	  \faGlobe & CookieLaw                 & 1  & -  &   & \faArrowCircleODown & Nessus                & 11 & -  &   & \faGlobe & Sucuri Site Check     & 3  & -  &   \\
	  \faGlobe & CookieMetrix              & 2  & 1  & \checkmark{} & \faGlobe & Nikto Online          & 2  & 2  & \checkmark{} & \faGlobe & SureOak               & 9  & 8  & \checkmark{}\\
	  \faGlobe & DNS Checker               & 1  & 1  & \checkmark{} & \faArrowCircleODown & \nmap{}                  & 14 & -  &   & \faGlobe & TheSEOTools           & 1  & 1  & \checkmark{}\\
	  \faGlobe & DNSTools                  & -  & -  &   & \faGlobe & \nmap{} Online           & 12 & 1  & \checkmark{}& \faGlobe & Tutorialspots         & 13 & 13 & \checkmark{}\\
	  \faGlobe & Dupli Checker             & 1  & -  & \checkmark{} & \faGlobe & Online SEO Tools      & 12 & 12 & \checkmark{} & \faGlobe & Url X-Ray             & 1  &  -  &   \\
	  \faGlobe & evilacid.com              & 12 & 12 & \checkmark{} & \faArrowCircleODown & OpenVAS               & 3  & -  &   & \faGlobe & Urlcheckr             & 10 & -  &   \\
	  \faGlobe & expandUrl                 & 1  & -  &   & \faArrowCircleODown & OWASP ZAP             & 4  & -  &   & \faGlobe & Urlex                 & -  & -  &   \\
	  \faGlobe & FreeDirectoryWebsites     & 13 & 13 & \checkmark{} & \faGlobe & Pentest-Tools         & 2  & 1  & \checkmark{} & \faGlobe & w-e-b.site            & 13 & 13 & \checkmark{}\\
	  \faGlobe & GDPR Cookie Scan          & -  & -  &   & \faGlobe & Port Checker          & 10 & -  &   & \faGlobe & W3dt.Net              & 12 & 11 & \checkmark{}\\
	  \faGlobe & GeekFlare                 & 12 & -  &   & \faGlobe & Redirect Check        & 11 & 10 & \checkmark{} & \faGlobe & Web Port Scanner      & -  & -  &   \\
	  \faGlobe & Hacker Target             & 13 & -  &   & \faGlobe & Redirect Detective    & 2  & -  & \checkmark{} & \faGlobe & Web Sniffer           & 14 & -  &   \\
	  \faGlobe & HTTP Tools                & 12 & 12 & \checkmark{} & \faGlobe & ReqBin                & 13 & -  &   & \faGlobe & WebConfs              & 13 & 12 & \checkmark{}\\
	  \faGlobe & httpstatus.io             & 14 & -  &   & \faGlobe & Resplace              & 12 & -  &   & \faArrowCircleODown & WebMap                & 14 & 1  & \checkmark{}\\
	  \faArrowCircleODown & InsightVM                 & 3  &  -  &   & \faGlobe & RexSwain.com          & 13 & 1  & \checkmark{}& \faGlobe & What Is My IP         & 12 & -  &   \\
	  \faGlobe & InternetMarketingNinjas & 1  & -  &   & \faGlobe & Search Engine Reports & 1  & 1  & \checkmark{} & \faArrowCircleODown & WMap                  & 12 & 10 & \checkmark{} \\
		\hline
	\end{tabularx}
\end{table*}

In Table~\ref{tab:results}, for each \scantool{} we report the number of
tainted and vulnerable flows (\textbf{T} and \textbf{V}, respectively) detected by \toolname{}.
After running \toolname{}, we also conducted a manual vulnerability assessment of each \scantool{}. 
Under column \textbf{M}, \checkmark{} indicates that an XSS vulnerability was found by a human analyst starting from the outcome of \toolname{}.
It is worth noticing that only in one case, i.e., DupliChecker, \toolname{} resulted in a false negative w.r.t. the manual analysis.
By investigating the causes, we discovered that DupliChecker performs URL encoding on the tainted locations. This encoding, among other operations,
replaces white spaces with \verb|%20|, thus invalidating our payloads.
To effectively bypass URL encoding, we replaced white spaces (U+0020) with non-breaking spaces (U+00A0) that are not modified.
Thus, we defined a new polyglot payload that uses non-breaking spaces and we added it to the injection list included in \toolname{}.
Using this new payload, \toolname{} could also detect
the vulnerability in DupliChecker.

At the time of writing, all the vulnerabilities detected by \toolname{} have been reported to the tool vendors and
are undergoing a responsible disclosure process (see Appendix~\ref{sec:disclosure}).

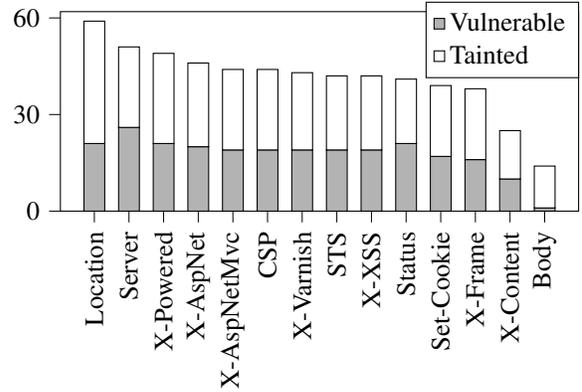
\begin{figure}
\input{figs/tainted-vulnerable-ordered.tex}
\caption{Frequency of tainted and vulnerable flows.}\label{fig:tainted-vulnerable-number}
\end{figure}

In Figure~\ref{fig:tainted-vulnerable-number} we show the frequency of the tainted and vulnerable flows over the 14 fields considered by \toolname{}.
Location has 59 tainted flows, the highest number, and 21 vulnerable flows.
Server only has 51 tainted flows, but it has 26 vulnerable flows, the highest number.
On the other hand, Body has only 14 tainted flows and only 1 vulnerable flow.
This highlights that most \scantool{s} sanitize the Body field in their reports.
The reason is that HTTP responses most likely contain HTML code in their Body.
Thus, sanitization is mandatory to preserve the report layout.
Also, the Body field is often omitted by the considered \scantool{s}.

\begin{figure}
	\centering
	\includegraphics[width=\columnwidth]{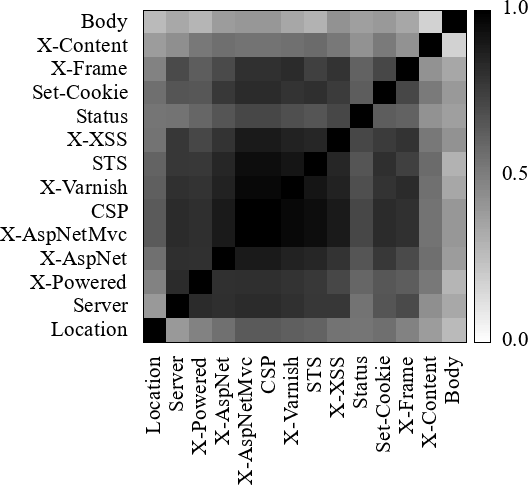}
	\caption{Correlation between tainted fields.}\label{fig:corr-taint}
\end{figure}

\begin{figure}
	\centering
	\includegraphics[width=\columnwidth]{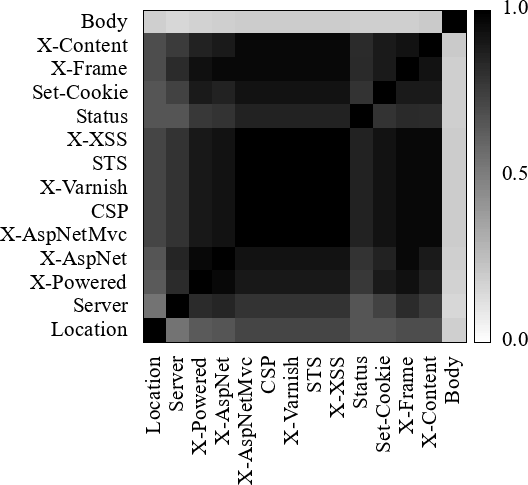}
	\caption{Correlation between vulnerable fields.}\label{fig:corr-vuln}
\end{figure}

In Figure~\ref{fig:corr-taint} and Figure~\ref{fig:corr-vuln} we show the correlation matrices for tainted and vulnerable fields, respectively.
From these matrices we observe a few, relevant facts.
We briefly discuss them below.

The first observation is that the Body field is almost unrelated to the other fields, both in terms of tainted and vulnerable flows.
This is somehow expected since the Body field is often neglected as discussed above.

Also the Location field is weakly correlated with the other fields.
This is due to the behavior of redirect checkers.
As a matter of fact, this category of \scantool{s} focus on Location, and, in most cases, ignore the other fields.
An in-depth evaluation of the behavior of redirect checkers is given in the application scenario of Section~\ref{sec:phishing}.

An argument similar to the previous one for Location also applies to Status Message.
The Status Message is typically used by \scantool{s} that carry out availability checks, e.g., to verify that a web site is up and running.

Finally, for what concerns all the other fields, we observe an extremely strong correlation.
The confirms the proposition of~\cite{lavrenovs2018securityheaders} about the security relevance of the headers that we are considering.
Indeed, most of the \scantool{s} included in our experiments report them all. 
This also highlights that the exposure of the \scantool{s} is not field-dependent, e.g., when a \scantool{} is vulnerable via one these fields, most likely it is also vulnerable via the others.



%% file: figs/tainted-vulnerable-ordered.tex
\begin{tikzpicture}

\definecolor{color1}{rgb}{1, 1, 1}
\definecolor{color0}{rgb}{0.7, 0.7, 0.7}

\begin{axis}[
legend cell align={left},
legend style={at={(1.001,1.05)}, fill opacity=1, draw opacity=1, text opacity=1, draw=black},
tick align=outside,
tick pos=left,
x grid style={white!69.0196078431373!black},
xmin=-0.98, xmax=13.98,
xtick style={color=black},
xtick={0,1,2,3,4,5,6,7,8,9,10,11,12,13},
ytick={0,30,60},
xticklabel style = {rotate=90.0},
xticklabels={Location, Server,X-Powered,X-AspNet,X-AspNetMvc,CSP,X-Varnish,STS,X-XSS,Status,Set-Cookie,X-Frame,X-Content,Body},
y grid style={white!69.0196078431373!black},
ymin=0, ymax=61.95,
height=0.5\columnwidth, width=\columnwidth,
ytick style={color=black}
]
\draw[draw=black,fill=color0] (axis cs:-0.3,0) rectangle (axis cs:0.3,21);
\draw[draw=black,fill=color0] (axis cs:0.7,0) rectangle (axis cs:1.3,26);
\draw[draw=black,fill=color0] (axis cs:1.7,0) rectangle (axis cs:2.3,21);
\draw[draw=black,fill=color0] (axis cs:2.7,0) rectangle (axis cs:3.3,20);
\draw[draw=black,fill=color0] (axis cs:3.7,0) rectangle (axis cs:4.3,19);
\draw[draw=black,fill=color0] (axis cs:4.7,0) rectangle (axis cs:5.3,19);
\draw[draw=black,fill=color0] (axis cs:5.7,0) rectangle (axis cs:6.3,19);
\draw[draw=black,fill=color0] (axis cs:6.7,0) rectangle (axis cs:7.3,19);
\draw[draw=black,fill=color0] (axis cs:7.7,0) rectangle (axis cs:8.3,19);
\draw[draw=black,fill=color0] (axis cs:8.7,0) rectangle (axis cs:9.3,21);
\draw[draw=black,fill=color0] (axis cs:9.7,0) rectangle (axis cs:10.3,17);
\draw[draw=black,fill=color0] (axis cs:10.7,0) rectangle (axis cs:11.3,16);
\draw[draw=black,fill=color0] (axis cs:11.7,0) rectangle (axis cs:12.3,10);
\draw[draw=black,fill=color0] (axis cs:12.7,0) rectangle (axis cs:13.3,1);
\draw[draw=black,fill=color1] (axis cs:-0.3,21) rectangle (axis cs:0.3,59);
\draw[draw=black,fill=color1] (axis cs:0.7,26) rectangle (axis cs:1.3,51);
\draw[draw=black,fill=color1] (axis cs:1.7,21) rectangle (axis cs:2.3,49);
\draw[draw=black,fill=color1] (axis cs:2.7,20) rectangle (axis cs:3.3,46);
\draw[draw=black,fill=color1] (axis cs:3.7,19) rectangle (axis cs:4.3,44);
\draw[draw=black,fill=color1] (axis cs:4.7,19) rectangle (axis cs:5.3,44);
\draw[draw=black,fill=color1] (axis cs:5.7,19) rectangle (axis cs:6.3,43);
\draw[draw=black,fill=color1] (axis cs:6.7,19) rectangle (axis cs:7.3,42);
\draw[draw=black,fill=color1] (axis cs:7.7,19) rectangle (axis cs:8.3,42);
\draw[draw=black,fill=color1] (axis cs:8.7,21) rectangle (axis cs:9.3,41);
\draw[draw=black,fill=color1] (axis cs:9.7,17) rectangle (axis cs:10.3,39);
\draw[draw=black,fill=color1] (axis cs:10.7,16) rectangle (axis cs:11.3,38);
\draw[draw=black,fill=color1] (axis cs:11.7,10) rectangle (axis cs:12.3,25);
\draw[draw=black,fill=color1] (axis cs:12.7,1) rectangle (axis cs:13.3,14);
\addlegendimage{only marks, mark=square*, color=color0, draw=black}
\addlegendentry{Vulnerable}
\addlegendimage{only marks, mark=square*, color=color1, draw=black}
\addlegendentry{Tainted}
\end{axis}

\end{tikzpicture}

%% file: case-study.tex
\section{Application Scenarios}\label{sec:case-study}

In this section, we present three application scenarios for our methodology.
For each scenario, we highlight the subclass of vulnerable \scantool{s},
the vulnerability and its impact if an attacker were to use it in the wild.
For each subclass of \scantool{s}, we chose a representative that we present as a concrete case study:
\nmap{} Online for as-a-service \scantool{s}, Metasploit Pro for on-premise ones, and CheckShortURL~\cite{checkshorturl} for redirect checkers.

\subsection{Scan Attribution}
\label{sec:fingerprint}

Attack attribution is a hot topic since it is often difficult or even impossible to achieve.
The main reasons are the structure of the network and some state-of-the-art technologies that enable clients anonymity.
For instance, analysts can use proxies, virtual private networks, and onion routing to hide the actual source of the requests from the recipient.
However, an injected browser may be forced to send identifying data directly inside the HTTP requests, so making network-level anonymization techniques ineffective.
In this section, we show how to attribute scans using our attacker model through an application scenario based on Nmap Online~\cite{nmap-online}.

\paragraph{\nmap{} Online vulnerability}\label{sec:nmap-online}
\nmap{} Online is a web application providing some of the functionalities of \nmap{}.
Users can scan a target with \nmap{} without having to install it on their machine.
Furthermore, since requests originate from the \nmap{} Online server, users can stay anonymous w.r.t. the scan target.
When users start a scan, they select the target IP and the scan type.
The \nmap{} Online website scans its target and displays the retrieved information to the user, e.g., server type and version.

\begin{figure}[t]
	\includegraphics[width=\columnwidth]{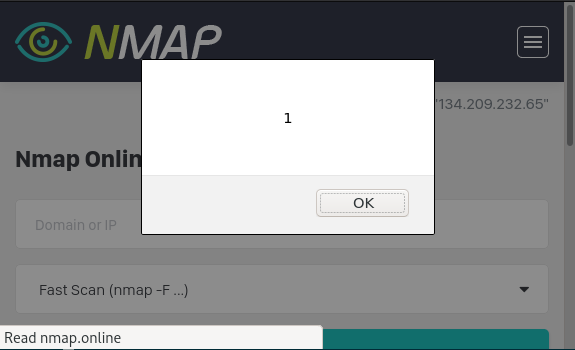}
    \caption{XSS PoC on Nmap Online.}
    \label{fig:nmap-online}
\end{figure}

\nmap{} Online reports suffered from an XSS vulnerability.\footnote{The vulnerability was fixed on March 24, 2020.}
Figure~\ref{fig:nmap-online} shows an injected report.
The injection occurs on the Server response header.
In this case, the Server field was set to \verb|<script>alert(1)</script>|.

\paragraph{Browser hooking}\label{sec:browser-hooking}

Since there is no guarantee that more than one scan will occur, we recur to browser hooking, which can be obtained with a single XSS payload.
A hooked browser becomes the client in a command and control (C2) infrastructure, thus actively querying the C2 server for instructions.
This allows the attacker to submit arbitrary commands afterward even when no other scans occur.

An effective way to achieve browser hooking is through BeEF~\cite{beef}.
In particular, the BeEF C2 client is injected via the script \emph{hook.js}.
For instance, we can deploy hook.js by setting the Server header to
\verb|<script src='http://[C2]/hook.js'></script>| where \verb|[C2]| is the IP address of the C2 server.

\paragraph{Fingerprinting}

The BeEF framework includes modules\footnote{\url{https://github.com/beefproject/beef/wiki/BeEF-modules}} for fingerprinting the victim host.
For instance, the \emph{browser} module allows us to get the browser name, version, visited domains, and even starting a video streaming from the webcam.
Similarly, the \emph{host} module allows us to retrieve data such as physical location and operating system details.
Some of these operations, e.g., browser fingerprinting, require no victim interaction.
Instead, others need the victim to take some actions, e.g., explicitly grant permission to use the webcam.
To overcome these hurdles, attackers usually employ auxiliary
techniques, e.g., credential theft, implemented by some BeEF modules,
e.g., \emph{social engineering}.
Finally, the overall fingerprinting process can be automated through the BeEF \emph{autorun rule engine}~\cite{beefARE}.

\subsection{Scanning host takeover}\label{sec:metasploit}

On-premise \scantool{s}, which run on the analyst's host, may have privileged, unrestricted access to the underlying platform.
In some cases, on-premise systems are provided with a user interface that includes both the reporting system and a control panel.
When such a user interface is browser-based, a malicious scan target can inject commands in the reporting system and perform lateral movements by triggering the \scantool{} controls.

The attack strategy abstractly described above must be implemented through concrete steps that are specific to the \scantool{}.
In this section, we show an implementation of this attack strategy for the popular \scantool{} Metasploit Pro.
In particular, we show how to perform lateral movements leading to a complete scanning host takeover through remote code execution (RCE).
Finally, we carry out an impact evaluation.

\paragraph{CVE-2020-7354 and CVE-2020-7355}
Metasploit Pro is a full-fledged penetration testing framework.
It has a browser-based UI that integrates both a scan reporting system and many controls for running the most common tasks, including host scanning.
Each command is executed by the Metasploit back-end, which is stimulated through a REST API. 

The vulnerability we found affects versions 4.17.0 and below.
It was remediated on May 14, 2020, with patch 4.17.1.\footnote{\url{https://help.rapid7.com/metasploit/release-notes/archive/2020/05/\#20200514}}
A malicious scan target can inject a Stored XSS payload in the UI.
Multiple pages are vulnerable, e.g., \verb|/hosts/:id| and \verb|/workspaces/:id/services|.

\begin{figure}[t]
	\includegraphics[width=\columnwidth]{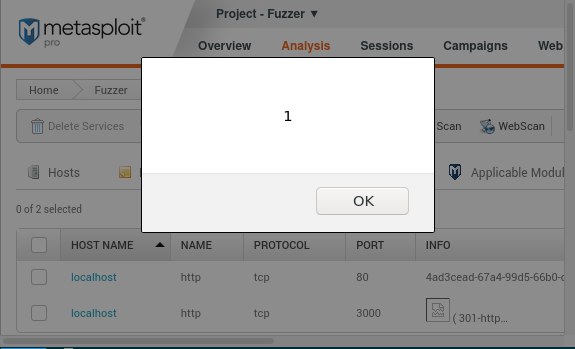}
	\caption{Stored XSS PoC on Metasploit Pro.}\label{fig:poc}
\end{figure}

Metasploit Pro fetches the Server header and displays it inside the \emph{INFO} column with no sanitization.
Figure~\ref{fig:poc} shows the effect of setting the Server header to \verb|<img src='x' onerror='alert(1)'/>| (as described in Section~\ref{sec:find-vuln-flows}).

\paragraph{Remote code execution}

We use the XSS vulnerability described above to gain a foothold in the browser on the scanning host.
For instance, we can inject a BeEF hook 
to remotely interact with the browser (as in Section~\ref{sec:browser-hooking}).
The hooked browser is the steppingstone to interact with the Metasploit Pro UI and trigger its controls.
Interestingly, Metasploit Pro includes a \emph{diagnostic console}, i.e., an embedded terminal that allows the analyst to run arbitrary commands on the underlying operating system.\footnote{\url{https://www.exploit-db.com/exploits/40415}}
Although the diagnostic console is disabled by default, the attacker can activate it through BeEF. In particular, the hooked browser is forced to perform a POST HTTP request to \verb|/settings/update_profile| with the parameter \verb|allow_console_access=1|.
Since the diagnostic console is a browser-embedded Metasploit terminal emulator, the attacker can submit commands from the BeEF interface.

\paragraph{Takeover impact}

The Metasploit Pro documentation\footnote{\url{https://metasploit.help.rapid7.com/docs/metasploit-web-interface-overview}} clearly states that ``Metasploit Pro Users Run as Root. If you log in to the Metasploit Pro Web UI, you can effectively run any command on the host machine as root''.
This opens a wide range of opportunities for the attacker.
Among them, the most impactful is to establish a \emph{reverse shell}.
The reasons are twofold.
First, opening a shell on the scanning host allows the attacker to execute commands directly on the operating system of the victim.
Thus, attacks are no longer tunneled through the initial vulnerability, which might become unavailable, e.g., if Metasploit Pro is terminated.
Second, a reverse shell works well even when certain network facilities, such as firewalls and NATs, are in place.
Indeed, although these facilities may prevent incoming connections, usually they allow outgoing ones.
Once a reverse shell is established, the attacker can access a permanent, privileged shell on the victim host.

\subsection{Enhanced phishing}\label{sec:phishing}

The goal of a phishing attack is to induce the
victim to commit a dangerous action, e.g., clicking an untrusted URL or opening an attachment.
In this section, we show how our attacker model changes phishing attacks, using CheckShortURL as an application scenario.

\paragraph{Traditional Phishing}

A common phishing scenario is that of an unsolicited email with a link pointing to a malicious web page, e.g., \verb|http://ev.il|.
The phishing site mimics a reputable, trusted web page.
For instance, the attacker may clone a bank's web site so that unaware users submit their access credentials.
Another technique is to provoke a reaction to an emotion, such as fear.
This happens, for instance, with menacing alerts about imminent account locking and malware infections.
Again, if victims believe that urgent action must be taken, they could overlook common precautions and, e.g., download dangerous files.

\paragraph{Defense mechanisms}

Most of the examples given above require the victim to open a phishing URL.
Common, unskilled users typically evaluate the trustworthiness of a URL by applying their common sense.\footnote{E.g., see \url{https://phishingquiz.withgoogle.com/}}
Nevertheless, techniques such as URL shortening and open redirects~\cite{shue08exploitable} masquerade the phishing URL to resemble a trusted domain.

Some online services may help the user to detect phishing attacks.
For instance, reputation systems and black/white lists, e.g., Web of Trust\footnote{\url{https://www.mywot.com}}, can be queried for a suspect URL.
However, phishing URLs often point to temporary websites that are unknown to these systems.

Since browsers automatically redirect without asking for confirmation, in~\cite{shue08exploitable} the authors highlight that victims can defend themselves by checking where the URL redirects without browsing it.
To this aim, several online services, e.g., CheckShortURL, do redirect checking to establish the final destination of a redirect chain.
Typically, the chain is printed in a report that the user inspects before deciding whether to proceed or not.

\paragraph{Exploiting redirect checkers}

Redirect locations are contained in the Location header of the HTTP response asking for a redirection.
According to our attacker model, this value is controller by the attacker.
Thus, if the victim uses a vulnerable redirect checker, the report may convey an attack to the user browser.
Since the goal is phishing, the attacker has two possibilities, i.e., forcing the URL redirection and exploit the \scantool{} reputation.

In the first case, the attacker delivers an XSS payload such as \verb|window.location = "http://ev.il/"|.
When it is executed, the browser is forced to open the given location and to redirect the user to the phishing site.

The second case is even more subtle.
Since the XSS attack is delivered by the \scantool{}, the attacker can perform a phishing operation and ascribe it to the reporting system.
For instance, the attacker can make the user browser download a malicious file pretending to be the \scantool{} pdf report. 
In this way, the attacker abuses the reputation of the \scantool{} to lure the victim.
This can be achieved with the following payload.

\noindent
\verb|window.location="http://tmpfiles.org/report.pdf"|

\begin{figure}[t]
    \includegraphics[width=\columnwidth]{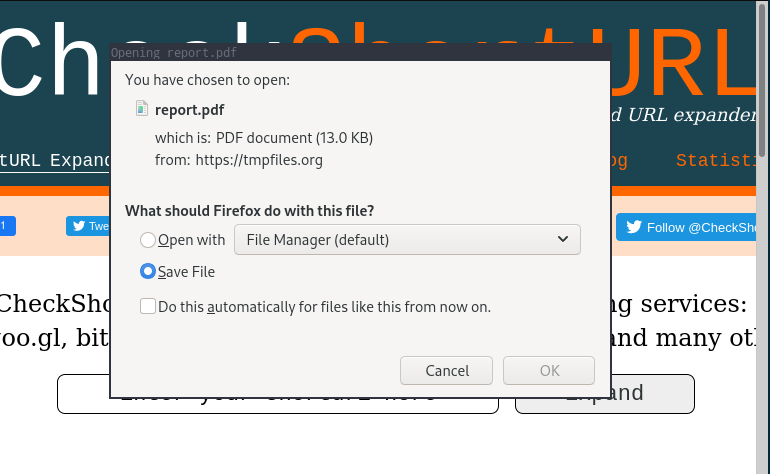}
	\caption{Phising through CheckShortURL.}
	\label{fig:checkshorturl}
\end{figure}

\noindent
The effect of injecting such a payload in CheckShortURL is shown in Figure~\ref{fig:checkshorturl}.

%% file: related.tex
\section{Related Work}\label{sec:related}

In this section, we survey the related literature.

\subsection{Attacking the attacker}

Although not frequent in the literature, the idea of attacking the attackers is not completely new.
Its common interpretation is that the victim of an attack carries out a counter-strike against the host of the aggressor.
However, even tracking an attack to its actual source is almost impossible if the attacker takes proper precautions (as discussed in Section~\ref{sec:fingerprint}).
To the best of our knowledge, we are the first to consider the response-based exploitation of the attackers \scantool{s}.

Djanali et al.~\cite{djanali2014aggressive} define a low-interaction honeypot that simulates vulnerabilities to lure the attackers to open a malicious website.
When this happens, the malicious website delivers a browser exploitation kit.
The exploitation relies on a LikeJacking~\cite{likejacking} attack to obtain information about the attacker's social media profile.
Unlike our approach, their proposal substantially relies on social engineering and does not consider vulnerabilities in the attacker's equipment.

Also, Sintov~\cite{sintsov2013honeypot} relies on a honeypot to implement a \emph{reverse penetration} process.
In particular, his honeypot attempts to collect data such as the IP address and the user agent of the attacker.
Again, this proposal amounts to retaliating against the attackers after identifying them.

In terms of vulnerabilities, some researchers already reported weaknesses in \scantool{s}.
The closest to our work is CVE-2019-5624~\cite{cve195624}, a vulnerability in RubyZip that also affects Metasploit Pro.
This vulnerability allows attackers to exploit \emph{path traversal} to create a \emph{cron job} that runs arbitrary code, e.g., to create a reverse shell.
To achieve this, the attacker must import a malicious file in Metasploit Pro as a new project.
However, as for~\cite{djanali2014aggressive}, this attack requires social engineering as well as other conditions (e.g., about the OS used by the attacker).
As far as we know, this is the only other RCE vulnerability reported for Metasploit Pro.
Instead, apart from ours, no XSS vulnerabilities have been reported.

\subsection{Security scanners assessment}

Several authors considered the assessment of security scanners.
However, they mainly focus on their effectiveness
and efficiency in detecting vulnerabilities.

Doupé et al.~\cite{doupe2010whyjohnny} present WackoPicko, an intentionally vulnerable web application designed to benchmark the effectiveness of security scanners.
The authors provide a comparison of how open source and commercial scanners perform on the different vulnerabilities contained in WackoPicko.

Holm et al.~\cite{holm2011quantitative} perform a quantitative evaluation of the accuracy of security scanners in detecting vulnerabilities.
Moreover, Holm~\cite{holm2012performance} evaluated the performance of network security scanners,
and the effectiveness of remediation guidelines.

Mburano et al.~\cite{mburano2018evalowasp} compare the performance of OWASP ZAP and Arachni.
Their tests are performed against the OWASP Benchmark Project~\cite{owasp-benchmark} and the Web Application 
Vulnerability Security Evaluation Project (WAVSEP)~\cite{wavsep}.
Both these projects aim to evaluate the accuracy, coverage,
and speed of vulnerability scanners.

To the best of our knowledge, there are no proposals
about the security assessment of \scantool{s}.
Among the papers listed above, none consider our attacker
model or, in general, the existence of security vulnerabilities in security scanners.

\subsection{Vulnerability detection}

Many authors proposed techniques to detect software
vulnerabilities.
In principle, some of these proposals can be applied
to \scantool{s}.

The general structure of vulnerability testing
environments was defined by Kals et al.~\cite{kals2006secubat}.
Our \tee{} implements their abstract framework by adapting it to inject responses instead of requests.
The main difference is our test stub, that receives the requests from the \scantool{} under test.
We substitute the crawling phase with tainted flow enumeration (see Section~\ref{sec:find-tainted-flows}).
During the attack phase, we substitute the payload list with a list of polyglots, which reduces testing time.
Our exploit checker implements their analysis module as
we also deal with XSS.

Many authors have proposed techniques to perform vulnerability detection through dynamic taint analysis.
For instance, Xu et al.~\cite{xu2005webdta} propose an approach that dynamically monitors sensitive sinks in PHP code. It rewrites PHP source code, injecting functions that monitor
data flows and detect injection attempts.

Avancini and Ceccato~\cite{avancini2010towardsdta} also use dynamic taint analysis to carry out vulnerability
detection in PHP applications.
Briefly, they implement a testing methodology
aiming at maximizing the code coverage.
To check whether a certain piece of code was
executed, they rewrite part of the application under test to deploy local checks.

These approaches rely on inspecting and manipulating the source code of the application under test.
Instead, we work under a black-box assumption.

Besides vulnerability detection,
some authors even use dynamic taint analysis
to implement exploit detection and prevention methodologies.
Vogt et al.~\cite{vogt07cross} prevent XSS attacks by combining dynamic and static taint analysis in a hybrid approach.
Similarly, Wang et al.~\cite{wang2018ttxss} detect DOM-XSS attacks using dynamic taint analysis.
Both these approaches identify sensitive data sinks
in the application code and monitor whether untrusted, user-provided input reaches them.

Dynamic taint analysis techniques were also proposed for
detecting vulnerabilities in binary code.

Newsome and Song~\cite{newsome05taintcheck} propose \emph{TaintCheck}, a methodology that leverages dynamic taint analysis to find attacks in commodity software.
\emph{TaintCheck} tracks tainted sinks and detects when an attack reaches them.
It requires a monitoring infrastructure to achieve this. 

Clause et al.~\cite{clause07dytan} propose a generic dynamic taint analysis framework.
Similarly to~\cite{newsome05taintcheck}, Clause et al. implement their technique for x86 binary executables.
However, the theoretical framework could be adapted to fit our methodology.

In principle, the exploit prevention techniques mentioned above might be used to mitigate some of the vulnerabilities
detected by \toolname{}.
However, they do not deal with vulnerability detection.
Moreover, they require access to the application code.


%% file: conclusion.tex
\section{Conclusion}
\label{sec:conclusion}

In this paper we introduced a new methodology, based on a novel attacker model, to detect vulnerabilities in \scantool{s}.
We implemented our methodology and we applied our prototype \toolname{} to \ntools{} real-world \scantool{s}.
Our experiments resulted in the discovery of \nvuln{}
new vulnerabilities.
These results confirm the effectiveness of our methodology
and the relevance of our attacker model.


%% file: disclosure.tex
\section{Vulnerability Disclosure}\label{sec:disclosure}

All the vulnerabilities reported in this paper were promptly notified to the \scantool{} vendors.
We based our responsible disclosure process on the ISO 29147\footnote{\url{https://www.iso.org/standard/72311.html}} guidelines.
Below, we describe each disclosure step in detail and the vendors feedback.

\subsection{First contact}

The first step of our responsible disclosure process consisted of a non-technical email notification to each vendor.
We report our email template below.

\begin{lstlisting}[basicstyle=\ttfamily, frame=single]
Dear <scanning system vendor>,

my name is <identification and links>

As part of my research activity on a 
novel threat model, I found that your
platform is most likely vulnerable 
to XSS attacks.
In particular, the vulnerability I
discovered might expose your end-users 
to concrete risks.

For these reasons, I am contacting 
you to start a responsible disclosure 
process. In this respect, I am kindly 
asking you to point me to the right 
channel (e.g., an official bug bounty 
program or a security officer to 
contact).

Kind regards
\end{lstlisting}

We sent the email through official channels, e.g., contact mail or form, when available.
For all the others, we tried with a list of 13 frequent email addresses, including
security@, webmaster@, contact@, info@, admin@, support@.

In 5 cases the previous attempts failed.
Thus, we submitted the corresponding vulnerabilities to OpenBugBounty.\footnote{\url{https://www.openbugbounty.org}}

\subsection{Technical disclosure}

After the vendor answered our initial notification, providing us with the technical point of contact, we sent a technical report describing the vulnerability.
The report was structured according to the following template, which was accompanied by a screenshot of the PoC exploit inside their system.

\begin{lstlisting}[basicstyle=\ttfamily, frame=single]
The issue is a Cross-Site Scripting 
attack on your online vulnerability 
scanning tool <scanning system name>.

This exposes your users to attacks, 
possibly leading to data leakage and 
account takeover.

A malicious server can answer with XSS 
payloads instead of its standard headers.
For example, it could answer with this 
(minimal) HTTP response:

<minimal PoC for the scanning system>

Since your website displays this data in 
a report, this code displays a popup on 
the user page, but an attacker can 
include any JavaScript code in it, 
taking control of the user browser (see 
https://beefproject.com/), and hence 
make them perform actions on your 
website or steal personal information.

I attached a screenshot of the PoC 
running on your page. The PoC is 
completely harmless, both for your 
website and for you to test.
I also hosted a malicious (but harmless) 
server here if you want to reproduce the 
issue: <test stub network address>

You can perform any scan you want 
against it (please let me know if it is 
offline).
\end{lstlisting}

In a few cases we extended the report with additional details, requested by some vendors.
For example, some of them asked for the CVSSv3\footnote{\url{https://nvd.nist.gov/vuln-metrics/cvss/v3-calculator}} calculation link
and an impact evaluation specifically referring their \scantool{}.

\subsection{Vendors feedback}

Out of the \nvuln{} notifications, we received 12 responses to the first contact message.
All the responses arrived within 2 days.
Among the notified vendors 5 fixed the vulnerability within 10 days.
Another vendor informed us that, although they patched their \scantool{}, they started a more general investigation of the vulnerability and our attacker model.
This will result in a major update in the next future.
Finally, after fixing the vulnerability, one of the vendors asked us not to appear in our research.